\documentclass[journal]{IEEEtran}
\IEEEoverridecommandlockouts
\usepackage{graphicx}
\usepackage{placeins}
\usepackage{float}
\usepackage{tabularx,colortbl}
\usepackage{amsthm}
\usepackage{amsmath}
\usepackage{mathtools}
\usepackage{amssymb}
\usepackage[named]{algo}
\usepackage[noend]{algpseudocode}
\usepackage{algorithm}
\usepackage{verbatim}
\usepackage{url}
\usepackage{cases}
\usepackage{amsmath}
\usepackage{amsfonts}
\usepackage{amssymb}
\usepackage{verbatim}
\usepackage{multirow}
\usepackage{graphicx}
\usepackage{epstopdf}
\usepackage{subfigure}
\usepackage{graphics}
\usepackage{color}
\usepackage[normalem]{ulem}  
\usepackage{cancel}
\usepackage{soul} 
\usepackage[bordercolor=gray!20,backgroundcolor=pink,linecolor=none
,textsize=scriptsize,textwidth=1.08cm]{todonotes}
\usepackage{siunitx}

\usepackage{tikz}

\IEEEoverridecommandlockouts

\begin{document}
\title{High Resolution FDMA MIMO Radar}
\author{David Cohen, Deborah Cohen, \emph{Student IEEE} and Yonina C. Eldar, \emph{Fellow IEEE} \thanks{
This project is funded by the European Union's Horizon 2020 research and innovation program under grant agreement No. 646804-ERC-COG-BNYQ, and by the Israel Science Foundation under Grant no. 335/14. Deborah Cohen is grateful to the Azrieli Foundation for the award of an Azrieli Fellowship. }}
\maketitle


\maketitle
\begin{abstract}
Traditional multiple input multiple output radars, which transmit orthogonal coded waveforms, suffer from range-azimuth resolution trade-off. In this work, we adopt a frequency division multiple access (FDMA) approach that breaks this conflict. We combine narrow individual bandwidth for high azimuth resolution and large overall total bandwidth for high range resolution. We process all channels jointly to overcome the FDMA range resolution limitation to a single bandwidth, and address range-azimuth coupling using a random array configuration.
\end{abstract}

    \section{Introduction}

Multiple input multiple output (MIMO) \cite{fishler2004mimo} radar combines several antenna elements both at the transmitter and receiver. Unlike phased-array systems, each transmitter radiates a different waveform, which offers more degrees of freedom \cite{li2009mimo}. Today, MIMO radars appear in many military and civilian applications including ground surveillance \cite{lesturgie2011some, martinez2007uwb}, automotive radar \cite{lutz201277, schuler2005array}, interferometry \cite{kim2007investigation}, maritime surveillance \cite{lesturgie2011some, anderson2010mimo}, through-the-wall radar imaging for urban sensing \cite{masbernat2010mimo} and medical imaging \cite{li2009mimo, pancera2010ultra}. There are two main configurations of MIMO radar, depending on the location of the transmitting and receiving elements; collocated MIMO \cite{li2007mimo} in which the elements are close to each other relatively to the working wavelength, and multistatic MIMO \cite{haimovich2008mimo} where they are widely separated. In this work, we focus on collocated MIMO systems.

MIMO radar presents significant potential for advancing state-of-the-art modern radar in terms of flexibility and performance. Collocated MIMO radar systems exploit the waveform diversity, based on mutual orthogonality of the transmitted signals \cite{li2009mimo}. Consequently, the performance of MIMO systems can be characterized by a virtual array corresponding to the convolution of the transmit and receive antenna locations. In principle, with the same number of antenna elements, this virtual array may be much larger and thus achieve higher resolution than an equivalent traditional phased array system~\cite{bliss2003multiple, rabideau2003ubiquitous, li2009diversity}. 

The orthogonality requirement, however, poses new theoretical and practical challenges. Choosing proper waveforms is a critical task for the implementation of practical MIMO radar systems. In addition to the general requirements on radar waveforms such as high range resolution and low sidelobes, MIMO radar waveforms must satisfy good orthogonality properties. In practice, it is difficult to find waveform families that perfectly satisfy all these demands \cite{cattenoz2015mimo}. Comprehensive evaluation and comparison of different types of MIMO radar waveforms is presented in \cite{gini2012waveform, rabaste2013signal, sun2014analysis}. The main waveform families considered are time, frequency and code division multiple access, abbreviated as TDMA, FDMA and CDMA, respectively. These may either be implemented in a single pulse, namely in the fast time domain, referred to as intra-pulse coding or in a pulse train, that is in the slow time domain, corresponding to inter-pulse coding. We focus on the former technique, which is most popular. More details on inter-pulse coding can be found in \cite{gini2012waveform, sun2014analysis}.

An intuitive and simple way to achieve orthogonality is using TDMA, where the transmit antennas are switched from pulse to pulse, so that there is no overlap between two transmissions \cite{ender2009system}. Since the transmission capabilities of the antennas are not fully utilized, this approach induces significant loss of transmit power \cite{gini2012waveform}, resulting in signal to noise ratio (SNR) decrease and much shorter target detection range. More efficient schemes have been proposed, such as circulating MIMO waveforms \cite{rabaste2013signal}. However, this technique suffers from loss in range resolution \cite{rabaste2013signal, sun2014analysis}.

Another way to achieve orthogonality of MIMO radar waveforms is FDMA, where the signals transmitted by different antennas are modulated onto different carrier frequencies. This approach suffers from several limitations. First, due to the linear relationship between the carrier frequency and the index of antenna element, a strong range azimuth coupling occurs when using the classic virtual uniform linear array (ULA) configuration \cite{cattenoz2015mimo, rabaste2013signal, sun2014analysis}. To resolve this aliasing, the authors in \cite{rias} use random carrier frequencies, which creates high sidelobe level. These may be mitigated by increasing the number of transmit antennas, which in turn increases system complexity. The second drawback of FDMA is that the range resolution is limited to a single waveform's bandwidth, rather than the overall transmit bandwidth \cite{stralka2011miso, vaidyanathan2009mimo}. To increase range resolution, the authors of \cite{donnet2006mimo, donnet2006combining} use an inter-pulse stepped frequency waveform (SFW), utilizing the total bandwidth over the slow time \cite{richards2005fundamentals, levanon2004radar}. However, SFW leads to range-Doppler coupling \cite{wehner1987high} and the pulse repetition frequency (PRF) increases proportionally to the number of steps increasing range ambiguities \mbox{\cite{ender2009system}, \cite{lord2000aspects}}.

In the popular CDMA approach, signals transmitted by different antennas are modulated using distinct series of orthogonal codes, so that they can be separated in the radar receiver. Although perfect orthogonality cannot in general be achieved, code families, such as Barker \cite{barker1953group}, Hadamard or Walsh \cite{tse2005fundamentals} and Gold \cite{gold1967optimal} sequences, present features close to orthogonality. CDMA requires good code design~\cite{li2009mimo} and may suffer from high range sidelobes depending on cross-correlation properties of the code sequence \cite{sun2014analysis}. More importantly, the narrowband assumption, that ensures constant delays over the channels, creates a trade-off between azimuth and range resolution, which can be a limiting factor for high resolution applications, by requiring either small aperture or small total bandwidth. In CDMA, the total bandwidth is equal to the individual bandwidth of each waveform, creating a conflict between large desired bandwidth for high range resolution and large virtual aperture for high azimuth resolution~\cite{vaidyanathan2008mimo}. The trade-off comes from the beamforming performance degradation when using wideband signals, since this operation is frequency dependent \hbox{\cite{ward1995theory}}. This dependency is quite severe in MIMO configurations where the virtual array is large. Several works \hbox{\cite{chou1995frequency}}, \hbox{\cite{liu2004new}} incorporate filter banks to ensure frequency invariance. However, in doing so, they increase system complexity at the receiver. In \hbox{\cite{vaidyanathan2008mimo}}, a smearing filter is adopted to address system complexity  which in turn leads to poor range resolution.

In this work, we adopt the FDMA approach and present an array design and processing method that overcome its drawbacks. First, to avoid range-azimuth coupling, we randomize the transmit and receive locations within the virtual array aperture. The idea of randomized frequencies has been used in single antenna radars \cite{wehner1987high} that employ SFW, to resolve range-Doppler coupling. There, hopped frequency sequences, namely with randomized steps for increasing the carriers, have been considered. Random arrays have been an object of research since the 1960s \hbox{\cite{lo1964mathematical}}. Recently, in \hbox{\cite{rossi2014spatial}}, the authors adopt random MIMO arrays to reduce the number of elements required for targets' detection using sub-Nyquist spatial sampling principles \cite{SamplingBook}. Here, we use a random array to deal with the coupling issue while keeping the number of elements as in traditional MIMO. We empirically found that randomizing the antenna locations rather than the frequencies exhibits better performance. 

Second, we process the samples from all channels jointly exploiting frequency diversity \cite{xu2015joint}, to overcome the range resolution limitation of a single bandwidth. A similar approach was used in \cite{wang2011space} in the context of MIMO synthetic aperture radar (SAR) with orthogonal frequency-division multiplexing linear frequency modulated (OFDM LFM) waveforms, where coherent processing over the channels allows to achieve range resolution corresponding to the total bandwidth. There, however, the MIMO array is composed of two uniform linear arrays (ULAs), both with spacing equal to half the wavelength. While avoiding range-elevation coupling, this approach yields poor elevation resolution \cite{li2009diversity}. Furthermore, in \cite{wang2011space}, the total bandwidth is limited by the narrowband assumption \mbox{\cite{vaidyanathan2008mimo}} and hence perpetuates the range-azimuth resolution conflict. Our approach does not require coding design and allows simpler matched filtering (MF) implementation than CDMA. 

The main contribution of this work is to show that using FDMA, the narrowband assumption may be relaxed to the individual bandwidth with appropriate signal processing. FDMA allows us to achieve a large overall received bandwidth over the channels while maintaining the narrowband assumption for each channel. This approach is inspired by SFW, first proposed in single antenna radars, in which a large overall bandwidth is achieved over the slow time to attain high range resolution while maintaining narrow instantaneous bandwidth. The range-azimuth resolution conflict may thus be solved by enabling large aperture for high azimuth resolution along with large total bandwidth for high range resolution. The narrowband assumption holds by requiring small individual bandwidth, breaking the traditional range-azimuth trade-off. In order to achieve range resolution corresponding to the overall bandwidth, we develop a recovery method that coherently processes all channels. This overcomes the traditional FDMA range resolution limitation to a single bandwidth.

We note that the radar cross section (RCS) may vary with frequency for distributed targets. This is beneficial in extended target applications, where orthogonal frequency division multiplexing (OFDM) may be used for additional frequency diversity as different scattering centers of a target resonate at different frequencies \cite{sen2011adaptive}. Unfortunately, when using coherent processing, the reflections from scatterers may interfere constructively or destructively depending on the signal frequency and the phases of the RCS for the individual scatterers \cite{skolnik, weinmann2010frequency}. In this work, we adopt the point-target assumption and perform coherent processing. Extended targets can then be modeled as the sum of point scatterers and high resolution may alleviate the above phenomena by separating the point scatterers over some resolution bins \cite{mensa1984wideband}.

This paper is organized as follows. In Section~\ref{sec:classic}, we review classic MIMO pulse-Doppler radar and processing. Our FDMA model is introduced in Section~\ref{sec:summer}, where range-azimuth coupling and beamforming are discussed. Section~\ref{sec:recovery} presents the proposed range-azimuth-Doppler recovery. Numerical experiments are presented in Section~\ref{sec:sim}, demonstrating the improved performance of our FDMA approach over classical CDMA.

\section{Classic MIMO Radar}
\label{sec:classic}

We begin by describing the classic MIMO radar architecture, in terms of array structure and waveforms, and the corresponding processing.

\subsection{MIMO Architecture}
\begin{figure}[!h]
\begin{center}
\includegraphics[width=0.5\textwidth]{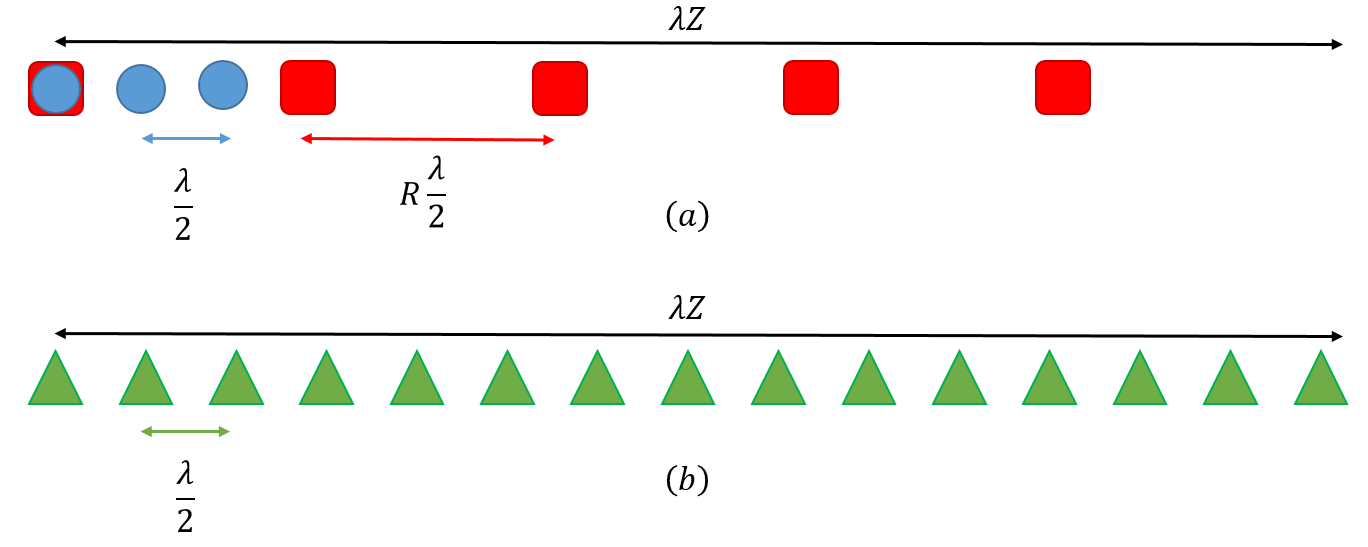}
\caption{Illustration of MIMO arrays: (a) standard array, (b) corresponding receiver virtual array.}
\label{fig:arrays1}
\end{center}
\end{figure}

The traditional approach to collocated MIMO adopts a virtual ULA structure \cite{chen2009signal}, where $R$ receivers, spaced by $\frac{\lambda }{2}$ and $T$ transmitters, spaced by $R\frac{\lambda }{2}$ (or vice versa), form two ULAs. Here, $\lambda$ is the signal wavelength. Coherent processing of the resulting $TR$ channels generates a virtual array equivalent to a phased array with $TR$ $\frac{\lambda }{2}$-spaced receivers and normalized aperture $Z=\frac{TR}{2}$. Denote by $\{\xi_{m}\}_{m=0}^{T-1}$ and $\{\zeta _{q}\}_{q=0}^{R-1}$ the transmitters and receivers' locations, respectively. For the traditional virtual ULA structure, $\zeta_q=\frac{q}{2}$ and $\xi_m=R\frac{m}{2}$. This standard array structure and the corresponding virtual array are illustrated in Fig.~\ref{fig:arrays1} for $R=3$ and $T=5$. The circles represent the receivers and the squares are the transmitters. In our work, we will consider a random array configuration \cite{rossi2014spatial}, where the antennas' locations are chosen uniformly at random within the aperture of the virtual array described above, that is 
$\{\xi_{m}\}_{m=0}^{M-1} \sim \mathcal{U} \left[ 0, Z \right]$ and $\{\zeta _{q}\}_{q=0}^{Q-1} \sim \mathcal{U} \left[ 0,Z \right]$, respectively. The corresponding virtual array has the same or a greater aperture than a traditional virtual array with the same number of elements, depending on the locations of the antennas at the far edges. The resulting azimuth resolution is thus at least as good as that of the traditional virtual ULA structure.

Each transmitting antenna sends $P$ pulses, such that the $m$th transmitted signal is given by 

	\begin{equation}\label{trMth}
  	s_m(t) = \sum_{p=0}^{P-1} {{{h}_{m}}\left( t -p\tau \right)}{{e}^{j2\pi {{f}_{c}}t}},\quad0\le t\le P\tau,
    \end{equation}
    where ${{h}_{m}}\left( t \right), 0\le m\le T-1 $ are orthogonal pulses with bandwidth $B_h$ and modulated with carrier frequency ${{f}_{c}}$. The coherent processing interval (CPI) is equal to $P \tau$, where $\tau$ denotes the pulse repetition interval (PRI). For convenience, we assume that $f_c \tau$ is an integer, so that the initial phase for every pulse $e^{-j2\pi f_c \tau p}$ is canceled in the modulation for $0 \leq p \leq P-1$ \cite{peebles2007radar}. The pulse time support is denoted by $T_p$, with $0 < T_p < \tau$.

MIMO radar architectures impose several requirements on the transmitted waveform family. Besides traditional demands from radar waveforms such as low sidelobes, MIMO transmit antennas rely on orthogonal waveforms. In addition, to avoid cross talk between the $T$ signals and form $TR$ channels, the orthogonality condition should be invariant to time shifts, that is $\int_{-\infty }^{\infty }{{{s}_{i}}\left( t \right)s_{j}^{*}\left( t-\tau_0  \right)dt}=\delta \left( i-j \right),$ for $i,j \in \left[0,T-1\right]$ and for all $\tau_0$. This property implies that the orthogonal signals cannot overlap in frequency (or time) \cite{vaidyanathan2008mimo}, leading to the FDMA (or TDMA) approach. Alternatively, time invariant orthogonality can be approximately achieved using CDMA.

Both FDMA and CDMA follow the general model \cite{cattenoz2013}:
\begin{equation} \label{eq:gen}
h_m(t)=\sum_{u=1}^{N_c} w_{mu} e^{j2\pi f_{mu}t} v(t-u\delta_t),
\end{equation}
where each pulse is decomposed into $N_c$ time slots with duration $\delta_t$. 
Here, $v(t)$ denotes the elementary waveform, $w_{mu}$ represents the code and $f_{mu}$ the frequency for the $m$th transmission and $u$th time slot. The general expression (\ref{eq:gen}) allows to analyze at the same time different waveform families. In particular, in CDMA, orthogonality is achieved by the code $\{w_{mu} \}_{u=1}^{N_c}$ and $f_{mu}=0$ for all $1 \leq u \leq N_c$. In FDMA, $N_c=1$, $w_{mu}=1$ and $\delta_t =0$. The center frequencies $f_{mu}=f_m$ are chosen in $[ -\frac{TB_h}{2}, \frac{TB_h}{2} ]$ so that the intervals $[{{f}_{m}}-\frac{{{B}_{h}}}{2},{{f}_{m}}+\frac{{{B}_{h}}}{2}]$ do not overlap. For simplicity of notation, $\{h_m(t)\}_{m=0}^{T-1}$ can be considered as frequency-shifted versions of a low-pass pulse $v(t)=h_0(t)$ whose Fourier transform $H_0\left( \omega  \right)$ has bandwidth $B_h$, such that
    \begin{equation}
    {{H}_{m}}\left( \omega  \right)={{H}_{0}}\left( \omega -2\pi {{f}_{m}} \right).
    \end{equation}
We adopt a unified notation for the total bandwidth $B_{\text{tot}}=TB_h$ for FDMA and $B_{\text{tot}}=B_h$ for CDMA.

Consider $L$ non-fluctuating point-targets, according to the Swerling-0 model \cite{skolnik}. 
Each target is identified by its parameters: radar cross section (RCS) $\tilde{\alpha}_l$, distance between the target and the array origin or range $R_l$, velocity $v_l$ and azimuth angle relative to the array $\theta_l$. Our goal is to recover the targets' delay $\tau_l=\frac{2R_l}{c}$, azimuth sine $\vartheta_l=\sin (\theta_l)$ and Doppler shift $f_l^D = \frac{2v_l}{c}f_c$ from the received signals. In the sequel, we use the terms range and delay interchangeably, as well as azimuth angle and sine, and velocity and Doppler frequency.

\subsection{Received Signal}

The transmitted pulses are reflected by the targets and collected at the receive antennas. The following assumptions are adopted on the array structure and targets' location and motion, leading to a simplified expression for the received signal.
\begin{itemize}
\item[\textbf{A1}] Collocated array - target RCS $\tilde{\alpha}_l$ and $\theta_l$ are constant over the array (see \cite{haim2006} for more details).
\item[\textbf{A2}] Far targets - target-radar distance is large compared to the distance change during the CPI, which allows for constant $\tilde{\alpha}_l$,
\begin{equation}
v_l P \tau \ll \frac{c \tau_l}{2}.
\end{equation}
\item[\textbf{A3}] Slow targets - low target velocity allows for constant $\tau_l$ during the CPI,
\begin{equation} \label{eq:A3_1}
\frac{2 v_l P \tau}{c} \ll \frac{1}{B_{\text{tot}}},
\end{equation}
and constant Doppler phase during pulse time $T_p$,
\begin{equation} \label{eq:A3_2}
f_l^D T_p \ll 1.
\end{equation}
\item[\textbf{A4}] Low acceleration - target velocity $v_l$ remains approximately constant during the CPI, allowing for constant Doppler shift $f^D_l$,
\begin{equation}
\dot{v_l} P \tau \ll \frac{c}{2f_cP\tau}.
\end{equation}
\item[\textbf{A5}] Narrowband waveform - small aperture allows $\tau_l$ to be constant over the channels,  
\begin{equation} \label{eq:A1}
\frac{2Z \lambda}{c} \ll \frac{1}{B_{\text{tot}}}.
\end{equation}
\end{itemize}

\begin{figure}[!h]
\begin{center}
\includegraphics[width=0.5\textwidth]{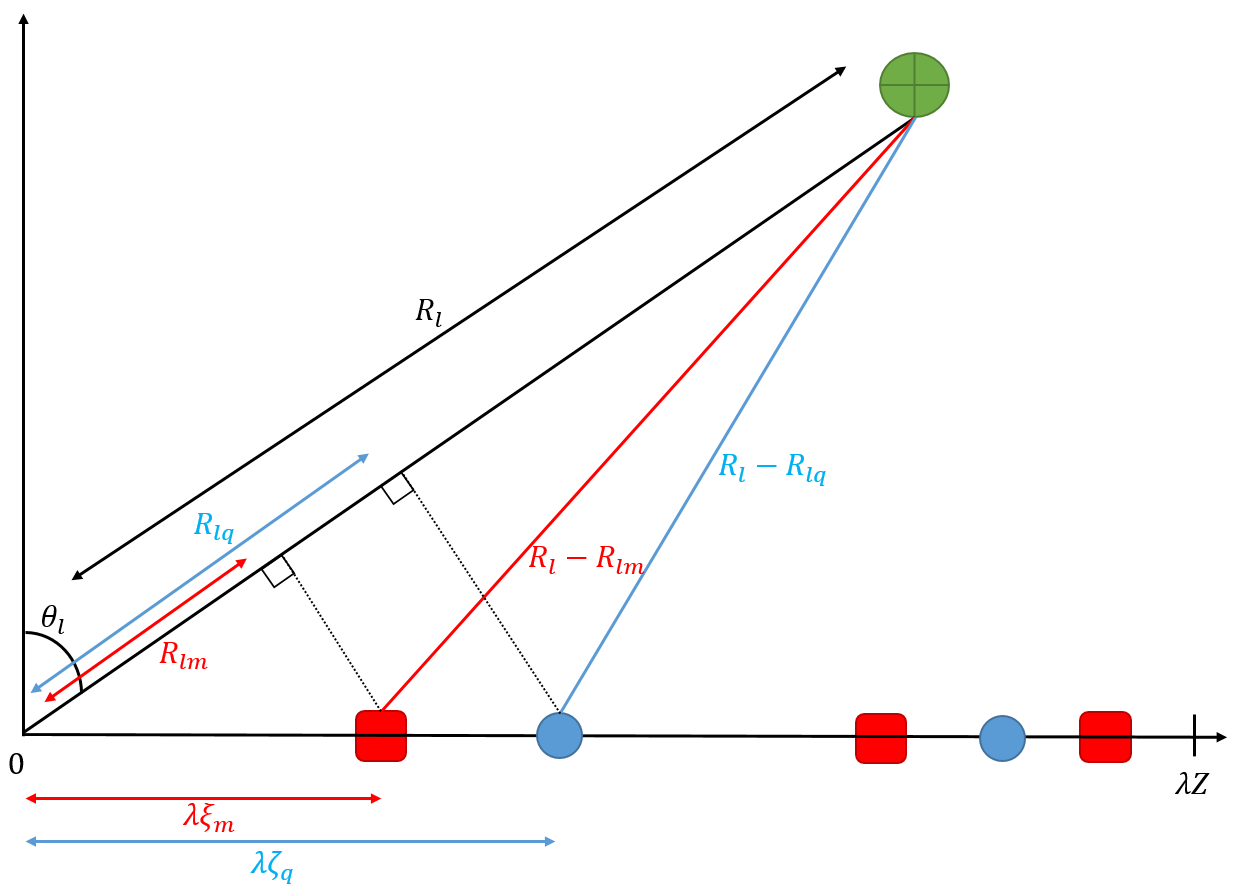}
\caption{MIMO array configuration.}
\label{fig:geom}
\end{center}
\end{figure}

Under assumptions \textbf{A1}, \textbf{A2} and \textbf{A4}, the received signal $\tilde{x}_q(t)$ at the $q$th antenna is a sum of time-delayed, scaled replica of the transmitted signals:
    \begin{equation}
    {\tilde{x}_{q}}\left( t \right) =  \sum\limits_{p-0}^{P-1} \sum\limits_{m=0}^{T-1} \sum\limits_{l=1}^{L} {\tilde{\alpha}_l {{s}_{m}}\left( \frac{c-v_l}{c+v_l} \left(t -\frac{R_{l,mq}}{c-v_l} \right)  \right)},
    \end{equation}
where $R_{l,mq}=2R_l-(R_{lm}+R_{lq})$, with $R_{lm}=\lambda \xi_m  \vartheta_l$ and $R_{lq}= \lambda \zeta_q \vartheta_l$ accounting for the array geometry, as illustrated in Fig.~\ref{fig:geom}.
The received signal can be simplified using assumptions \textbf{A3} and \textbf{A5}, as we now show.

We start with the envelope $h_m(t)$, and consider the $p$th frame and the $l$th target. From $c \pm v_l \approx c$, and neglecting the term $\frac{2v_lt}{c}$ using (\ref{eq:A3_1}), we obtain
\begin{equation} \label{eq:simp1}
h_m \left( \frac{c-v_l}{c+v_l} \left(t-\frac{R_{l,mq}}{c-v_l}  \right) -p\tau \right) = h_m(t-p\tau - \tau_{l,mq}).
\end{equation}
Here, $\tau_{l,mq}=\tau_l-\eta_{mq}\vartheta_l$ where $\tau_l=\frac{2R_l}{c}$ is the target delay, and $\eta_{mq}=(\xi_m + \zeta_q) \frac{\lambda}{c}$ follows from the respective locations between transmitter and receiver. We then add the modulation term of $s_m(t)$. Again using $c \pm v_l \approx c$, the remaining term is given by
\begin{equation} \label{eq:term2}
h_m(t-p\tau - \tau_{l,mq}) e^{j2\pi (f_c-f_l^D) \left(t-\tau_{l,mq} \right)}.
\end{equation}
After demodulation to baseband and using (\ref{eq:A3_2}), we further simplify (\ref{eq:term2}) to
\begin{equation} \label{eq:term3}
h_m(t-p\tau - \tau_{l,mq}) e^{-j2 \pi f_c \tau_l} e^{j2 \pi f_c \eta_{mq}\vartheta_l} e^{-j2 \pi f^D_l p\tau}.
\end{equation}
The three phase terms in (\ref{eq:term3}) correspond to the target delay, azimuth and Doppler frequency, respectively.
Last, from \textbf{A5}, the delay term $\eta_{mq}\vartheta_l$, that stems from the array geometry, is neglected in the envelope which becomes
\begin{equation} \label{eq:term_1}
 h_m(t-p\tau - \tau_l).
\end{equation}
Substituting (\ref{eq:term_1}) to (\ref{eq:term3}), the received signal at the $q$th antenna after demodulation to baseband is given by
 \begin{equation} \label{eq:rec_sig1}
 x_q \left( t \right) = \sum\limits_{p=0}^{P-1} \sum\limits_{m=0}^{T-1} \sum\limits_{l=1}^{L}  \alpha_l h_m \left( t- p\tau  - \tau _{l} \right) e^{j2 \pi f_c \eta_{mq}\vartheta_l} e^{-j2 \pi f^D_l p\tau},
\end{equation}
where $\alpha_l = \tilde{\alpha}_l e^{-j2 \pi f_c \tau_l}$.

\begin{figure}
\begin{center}
\includegraphics[width=0.5\textwidth]{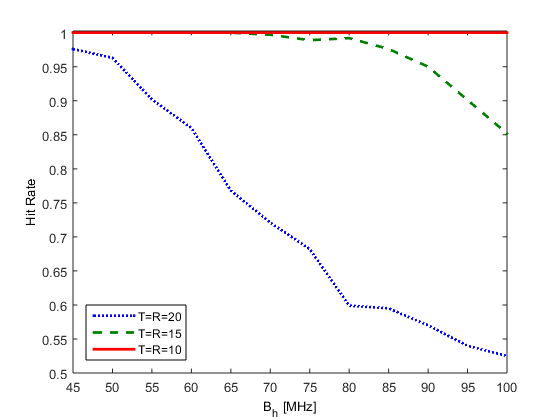}
\caption{Hit rate of MIMO radar with classical processing of CDMA waveforms with respect to  total bandwidth $B_{\text{tot}}=B_h$ and aperture $Z=TR/2$.}
\label{fig:cdma_prob}
\end{center}
\end{figure}

The narrowband assumption \textbf{A5} leads to a trade-off between azimuth and range resolution, by requiring either small aperture $Z$ or small total bandwidth $B_{\text{tot}}$, respectively. In CDMA, $B_{\text{tot}}=B_h$ so that \textbf{A5} limits the total bandwidth of the waveforms $h_m(t)$ \cite{vaidyanathan2008mimo}. This is illustrated in Fig.~\ref{fig:cdma_prob} which shows the performance of CDMA waveforms using the classical processing detailed below. We use bandlimited Gaussian pulses that are equivalent to CDMA, where each transmitter radiates $P=1$ pulse, and consider range-azimuth recovery in the absence of noise. We assume $L=5$ targets whose locations are generated uniformly at random, and adopt a hit-or-miss criterion as our performance metric. A ``hit" is defined as a range-azimuth estimate which is identical to the true target position up to one Nyquist bin (grid point) defined as $1/B_{\text{tot}}$ and $2/TR$ for the range and azimuth, respectively. Each experiment is repeated over 200 realizations. It can be seen that the recovery performance decreases with either increased bandwidth or aperture since in both cases \textbf{A5} does not hold. In the next section, we show that in FDMA, this assumption can be relaxed under appropriate processing so that the aperture is required to be smaller than the reciprocal of $B_h$ rather than $B_{\text{tot}}=TB_h$ as in (\ref{eq:A1}).


\subsection{Range-Azimuth-Doppler Recovery}
Classic collocated MIMO radar processing traditionally includes the following stages:

\begin{enumerate}
  \item \textbf{Sampling:} at each receiver, the signal $x_q(t)$ is sampled at its Nyquist rate $B_{\text{tot}}$.
  \item \textbf{Matched filter:} the sampled signal is convolved with a sampled version of $h_m(t)$, for $0 \leq m \leq T-1$. The time resolution attained in this step is $1/B_h$. In FDMA, this step leads to a limitation on the range resolution to a single channel bandwidth rather than the total bandwidth.
  \item \textbf{Beamforming:} correlations between the observation vectors from the previous step and steering vectors corresponding to each azimuth on the grid defined by the array aperture are computed. The spatial resolution attained in this step is $2/TR$. In FDMA, this stage leads to range-azimuth coupling, as illustrated in Section~\hbox{\ref{sec:coupl}}.
  \item \textbf{Doppler detection:} correlations between the resulting vectors and Doppler vectors, with Doppler frequencies lying on the grid defined by the number of pulses, are computed. The Doppler resolution is $1/P\tau$.
  \item \textbf{Peak detection:} a heuristic detection process is performed on the resulting range-azimuth-Doppler map. For example, the detection can follow a threshold approach \cite{richards2005fundamentals} or select the $L$ strongest point of the map, if the number of targets $L$ is known.
\end{enumerate}

\begin{figure}[!h]
\begin{center}
\includegraphics[width=0.5\textwidth]{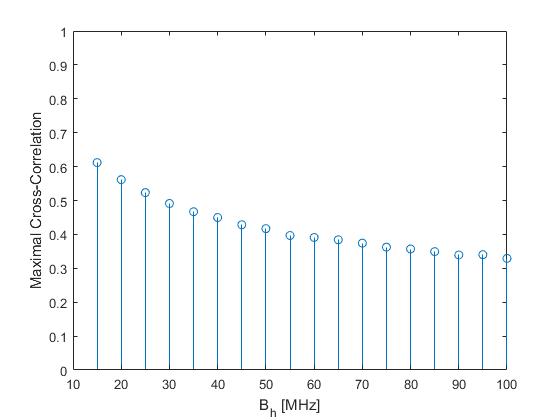}
\includegraphics[width=0.5\textwidth]{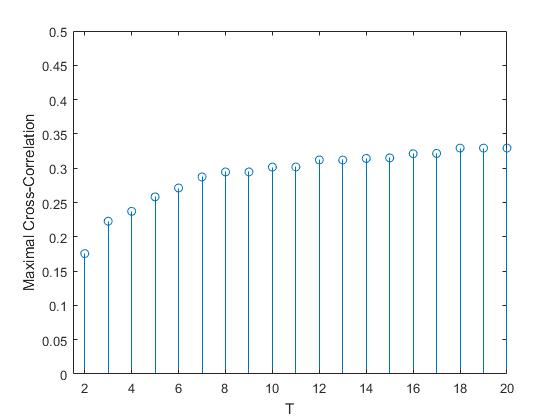}
\caption{Maximal cross-correlation of CDMA waveforms using $T_p=0.44 \mu$sec with respect to signal bandwidth $B_h$ with $T=20$ transmitters (top) and number of transmitters $T$ with bandwidth $B_h=100$MHz (bottom).}
\label{fig:sidelobes}
\end{center}
\end{figure}

CDMA is a popular MIMO approach even though it suffers from two main drawbacks. First, the narrowband assumption yields a trade-off between azimuth and range resolution. Second, achieving orthogonality through code design has proven to be a challenging task \cite{li2009mimo}. To illustrate this, we consider a set of orthogonal bandlimited Gaussian waveforms, generated using a random search for minimizing the cross-correlation between pairs of waveforms. That is, the set of waveforms is constructed so that to minimize the maximal cross-correlation between waveforms, through a non-exhaustive search. Figure~\ref{fig:sidelobes} shows the maximal cross-correlations between any pair of signals within the set. It can be seen that, when either the bandwidth $B_h$ is reduced or the number of transmit antennas increases, the maximal cross correlation of the CDMA waveforms increases. 

Meanwhile, classic FDMA has been almost neglected owing to its two main drawbacks. First, due to the linear relationship between the carrier frequency and the index of antenna element, a strong range-azimuth coupling occurs \cite{cattenoz2015mimo, rabaste2013signal, sun2014analysis}, as we illustrate in Section~\hbox{\ref{sec:coupl}}. To resolve this aliasing, the authors in \cite{rias} use random carrier frequencies, which creates high sidelobe levels. The second drawback of FDMA is that the range resolution is limited to a single waveform's bandwidth, namely $B_h$, rather than the overall transmit bandwidth $B_{\text{tot}}=TB_h$ \cite{stralka2011miso, vaidyanathan2009mimo}.

In the next section, we adopt the FDMA approach, in order to exploit the narrowband property of each individual channel to achieve both high range and azimuth resolution. To resolve the coupling issue, we randomly distribute the antennas, while keeping the carrier frequencies on a grid with spacing $B_h$. In the simulations, we show that random antenna locations yield smaller sidelobes than random carriers. Next, by processing the channels jointly, we achieve a range resolution of $1/B_{\text{tot}}=1/TB_h$ rather than $1/B_h$. This way, we exploit the overall received bandwidth that governs the range resolution, while maintaining the narrowband assumption for each channel, which is key to high azimuth resolution. In addition, no code design is required, which may be a challenging task \cite{li2009mimo}, as shown in Fig.~\ref{fig:sidelobes}. 

\section{FDMA System}
\label{sec:summer}

In this section, we describe our MIMO system based on joint channel processing of FDMA waveforms. Our FDMA processing differs from the classic CDMA approach introduced in Section \hbox{\ref{sec:classic}} in several aspects. First, the single channel processing, which is equivalent to matched filtering in step (2), is limited to $1/B_h$ whereas in FDMA we achieve resolution of $1/TB_h$. In addition, in FDMA the range depends on the channels while in CDMA, it is decoupled from the channels domain. Therefore, our processing involves range-azimuth beamforming while the classic approach for CDMA uses beamforming on the azimuth domain only as in step (3). The range dependency on the channels in FDMA is exploited to enhance the poor range resolution of the single channel $1/B_h$ to $1/TB_h = 1/B_{\text{tot}}$, as explained in the remainder of this section. Finally, combining the use of FDMA waveforms with our proposed processing reconciles the narrowband assumption with large total bandwidth for range resolution, enhancing range-azimuth resolution capabilities.

\subsection{Received Signal Model}

Our processing, described in Section \ref{sec:algo}, allows to soften the strict neglect of the delay term in the transition from (\ref{eq:term3}) to (\ref{eq:term_1}). We only remove $\eta_{mq}\vartheta_l$, that stems from the array geometry, from the envelope $h_0(t)$ rather than $h_m(t)$. Then, (\ref{eq:term_1}) becomes
\begin{equation} \label{eq:term1_1}
 h_m(t-p\tau - \tau_l)e^{j2\pi f_m \eta_{mq} \vartheta_l}.
\end{equation}
Here, the restrictive assumption \textbf{A5} is relaxed to $\frac{2Z \lambda}{c} \ll \frac{1}{B_h}$. We recall that, in CDMA, \textbf{A5} leads to a trade-off between azimuth and range resolution, by requiring either small aperture or small total bandwidth $B_{\text{tot}}$, respectively. Here, using the FDMA framework and the less rigid approximation (\ref{eq:term1_1}), we need only the single bandwidth $B_h$ to be narrow, rather than the total bandwidth $B_{\text{tot}}=TB_h$, eliminating the trade-off between range and azimuth resolution.

The received signal at the $q$th antenna after demodulation to baseband is in turn given by
 \begin{equation} \label{eq:rec_sig}
 x_q \left( t \right) = \sum\limits_{p=0}^{P-1} \sum\limits_{m=0}^{T-1} \sum\limits_{l=1}^{L}  \alpha_l h_m \left( t- p\tau  - \tau _{l} \right) e^{j2 \pi \beta_{mq} \vartheta _l} e^{-j 2\pi f^D_l p \tau},
\end{equation}
where $\beta_{mq} = \left( \zeta _{q}+\xi _m \right) \left( f_m \frac{\lambda}{c} +1 \right)$. 
In comparison with (\ref{eq:rec_sig1}), neglecting the delay term only in the narrowband envelope $h_0(t)$, (\ref{eq:rec_sig}) results in the extra term $e^{j2 \pi (\zeta _{q}+\xi _m) \vartheta _l f_m \lambda /c}$. This corrects the time of arrival differences  between channels, so that the narrowband assumption \textbf{A5} is required only on $h_0(t)$ with bandwidth $B_h$ and not on the entire bandwidth $B_{\text{tot}}$.
Intuitively, the waveforms are aligned to eliminate the arrival differences resulting from the array geometry with respect to $f_m$, thus enabling to detect the azimuth with respect to the central carrier $f_c$.

It will be convenient to express $x_q(t)$ as a sum of single frames
\begin{equation}
\label{eq:frames}
x_q(t)= \sum_{p=0}^{P-1} x_q^p(t),
\end{equation}
where
\begin{equation}
\label{eq:one_frame}
x_q^p(t)= \sum_{m=0}^{T-1} \sum_{l=1}^{L} \alpha_l h_m(t-\tau_l - p\tau) e^{j2 \pi \beta_{mq} \vartheta _l} e^{-j 2\pi f^D_l p \tau}.
\end{equation}
Our goal is to estimate the targets range, azimuth and velocity, i.e. to estimate ${{\tau}_{l}}$, ${{\vartheta}_{l}}$ and $f^D_l$ from $x_q(t)$.

\subsection{Frequency Domain Analysis}
\label{sec:fourier_coeff}

We begin by deriving an expression for the Fourier coefficients of the received signal, and show how the unknown parameters, namely $\tau_l$, $\vartheta_l$ and $f^D_l$ are embodied in these coefficients. We next turn to range-azimuth beamforming and its underlying resolution capabilities and discuss the range-azimuth coupling. Finally, we present our proposed recovery algorithm, which is based on FDMA waveforms. To introduce our processing, we start with the special case of $P=1$, namely a single pulse is transmitted by each transmit antenna. We show how the range-azimuth map can be recovered from the Fourier coefficients in time and space. Subsequently, we treat the general case where a train of $P>1$ pulses is transmitted by each antenna, and present a joint range-azimuth-Doppler recovery algorithm from the  Fourier coefficients.

The $p$th frame of the received signal at the $q$th antenna, namely $x_q^p(t)$, is limited to $t\in \left[ p\tau, (p+1)\tau \right]$ and thus can be represented by its Fourier series
    \begin{equation}
    \begin{array}{lll}
    x_q^p(t)=\sum\limits_{k\in \mathbb{Z}}{{{c}_{q}^p}\left[ k \right]{{e}^{j2\pi kt/ \tau}}},\quad t\in \left[ p\tau,(p+1)\tau \right],
    \end{array}
    \end{equation}
    where, for $-\frac{NT}{2} \leq k \leq \frac{NT}{2}-1$, with $N=\tau B_h$,
    \begin{equation}
    \label{eq:coeffq2}
    c_q^p\left[ k \right] = \frac{1}{\tau}\sum\limits_{m=0}^{T-1} \sum\limits_{l=1}^L \alpha_l e^{j2\pi \beta_{mq} \vartheta _l} e^{-j\frac{2\pi}{\tau}k \tau_l} e^{-j 2 \pi f^D_l p \tau} H_m \left(\frac{2 \pi}{\tau}k \right).
    \end{equation}    
    
Once the Fourier coefficients $c_{q}^p[k]$ are computed, we separate them into channels for each transmitter, by exploiting the fact that they do not overlap in frequency. Applying a matched filter, we have
   \begin{eqnarray}
    & & \tilde{c}_{q,m}^p  \left[ k \right]  =  c_q^p\left[ k \right]H_{m}^{*}\left( \frac{2\pi }{\tau}k \right)  \\
   & & =\frac{1}{\tau} \left| H_m\left( \frac{2\pi }{\tau}k \right) \right|^2  \sum\limits_{l=1}^{L}\alpha_l e^{j2 \pi \beta_{mq} \vartheta_l} e^{-j\frac{2\pi}{\tau} k \tau_l}e^{-j 2 \pi f^D_l p \tau}.  \nonumber
    \end{eqnarray}
Let ${{y}_{m,q}^p}\left[ k \right]=\frac{\tau}{|H_0\left( \frac{2\pi }{\tau}k \right)|^2}{{\tilde{c}}_{q,m}^p}\left[ k+{{f}_{m}}\tau \right]$ be the normalized and aligned Fourier coefficients of the channel between the $m$th transmitter and $q$th receiver. Then,
      \begin{equation}
    \label{coffAlligned2}
   y_{m,q}^p[k]=   \sum_{l=1}^{L} \alpha_l e^{j2\pi \beta_{mq} \vartheta_l}e^{-j\frac{2\pi }{\tau}k\tau_l} e^{-j2\pi f_m \tau_l} e^{-j 2 \pi f^D_l p \tau},
       \end{equation}
       for $-\frac{N}{2} \leq k \leq \frac{N}{2}-1$. 
Our goal then is to recover the targets' parameters $\tau_l$, $\theta_l$ $\vartheta_l$ and $f^D_l$ from $ y_{m,q}^p[k]$.

\subsection{Range-Azimuth Beamforming}
\label{sec:res_cpbl}
Let us now pause to discuss the range and azimuth resolution capabilities of the described model and processing as well as the coupling issue between the two parameters.
Since the Doppler frequency is decoupled from the range-azimuth domain, we assume that $P=1$ for the sake of clarity. Then, \hbox{(\ref{coffAlligned2})} can be simplified to 
\begin{equation}
\label{coffAlligned2_simp}
y_{m,q}[k]=   \sum_{l=1}^{L} \alpha_l e^{j2\pi \beta_{mq} \vartheta_l}e^{-j\frac{2\pi }{\tau}k\tau_l} e^{-j2\pi f_m \tau_l},
\end{equation} 
for $-\frac{N}{2} \leq k \leq \frac{N}{2}-1$. The azimuth is embodied in the first term and its resolution is related to the virtual array geometry governed by $\beta_{mq}$ as discussed in \mbox{\cite{li2009diversity}}. The delay is embodied in the second and third terms, which allows both high range resolution and large unambiguous range. The second term leads to a poor range resolution of $1/B_h$ corresponding to a single channel, while the resolution induced by the third term, which measures the effect of the delay on the transmit carrier, is dictated by the total bandwidth, namely $1/TB_h$. On the other hand, since $f_m$ is a multiple of $B_h$ in our configuration, the last term is periodic in $\tau_l$ with a limited period of $1/B_h$, whereas the second term is periodic in $\tau_l$ with period $\tau$ so that the corresponding unambiguous range is $c\tau /2$. Therefore, by jointly processing both terms we overcome the resolution and ambiguous range limitations and thus achieve a range resolution of $1/TB_h$ with unambiguous range of $\tau$, as summarized in Table \hbox{\ref{tab:a}}. 

\begin{figure}[!h]
\begin{center}
\includegraphics[width=0.5\textwidth]{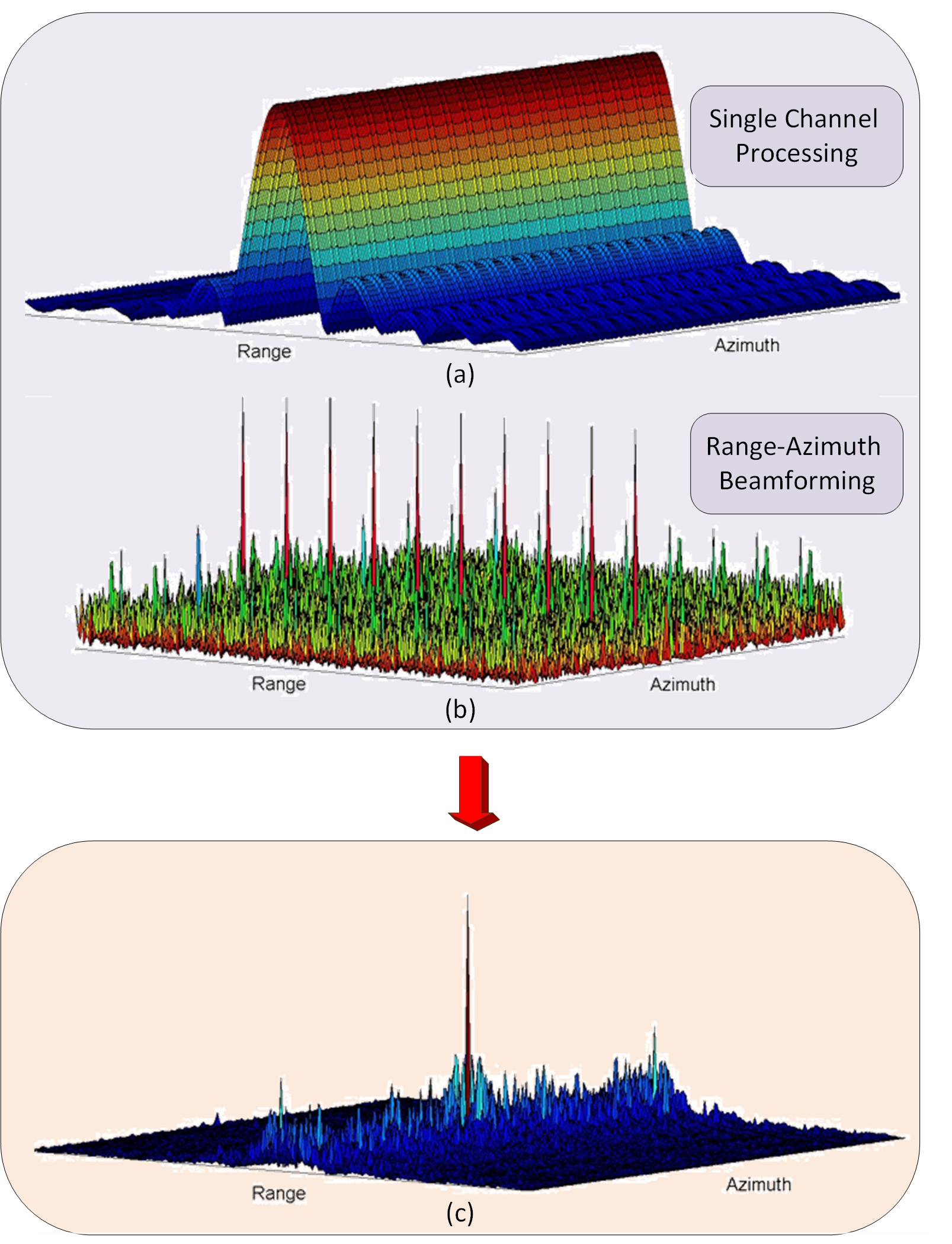}
\caption{Illustration of the resolution obtained by processing a single channel and by joint processing of all channels, using range-azimuth beamforming.}
\label{fig:illust}
\end{center}
\end{figure}

Both the first and third terms, which contain the azimuth and delay respectively, depend on the channels indexed by $m, q$ and thus need to be resolved simultaneously. This processing step is referred to as range-azimuth beamforming and will be discussed in the next section. The joint processing, which combines single-channel processing and range-azimuth beamforming, is illustrated in Fig.~\ref{fig:illust}. The poor range resolution that would be obtained by processing each channel separately can be seen in Fig.~\ref{fig:illust}(a). Range-azimuth beamforming, illustrated in Fig.~\ref{fig:illust}(b), achieves a higher resolution of $1/TB_h$, corresponding to the total bandwidth. However, the resulting range ambiguity can be clearly observed. Combining the single channel processing with range-azimuth beamforming yields joint range-azimuth recovery (Fig.~\ref{fig:illust}(c)). Note that the processing is not divided into these two steps, which are provided for illustration purposes only.

\begin{table}
\begin{center}
\caption{Range Resolution and Ambiguity}\label{tab:a}
\begin{tabular}
	{|>{\centering\arraybackslash} m{1.8cm}|>{\centering\arraybackslash} m{1.8cm}|>{\centering\arraybackslash} m{1.8cm}| @{}m{0pt}@{}}
	\hline
	\textbf{Term} & \textbf{Range Resolution} & \textbf{Unambiguous Range}  \\ 
	\hline
    $e^{-j\frac{2\pi }{\tau}k\tau_l}$ & \cellcolor{red!25} $1/B_h$  &  \cellcolor{green!25} $\tau$  &\\  [10pt]
    \hline
    $e^{-j2\pi f_m \tau_l}$ & \cellcolor{green!25}$1/TB_h$  & \cellcolor{red!25}$1/B_h$ &\\ [10pt]
    \hline
    \textbf{Joint Processing}& \cellcolor{green!25}$1/TB_h$  & \cellcolor{green!25}$\tau$  &\\  [10pt] 
    \hline
\end{tabular} 
\end{center}
\end{table}

\subsection{Range-Azimuth Coupling}
\label{sec:coupl}

As explained above, range-azimuth beamforming over the channels involves the estimation of both parameters from one dimension, the channel dimension. Therefore, it requires a one-to-one correspondence between the phases over the channels to range-azimuth pairs in order to prevent coupling. This ensures that each phase over the channels, expressed by the first and third terms of (\ref{coffAlligned2_simp}), corresponds to a unique azimuth-range pair. We next illustrate the range-azimuth coupling which occurs, in particular, in the case where the antennas are located according to the conventional virtual ULA structure shown in Fig.~\ref{fig:arrays1}, and the carrier frequencies are selected on a grid such that $f_m=(m-\frac{T-1}{2})B_h$. 

\begin{figure}
\begin{center}
\includegraphics[width=0.5\textwidth]{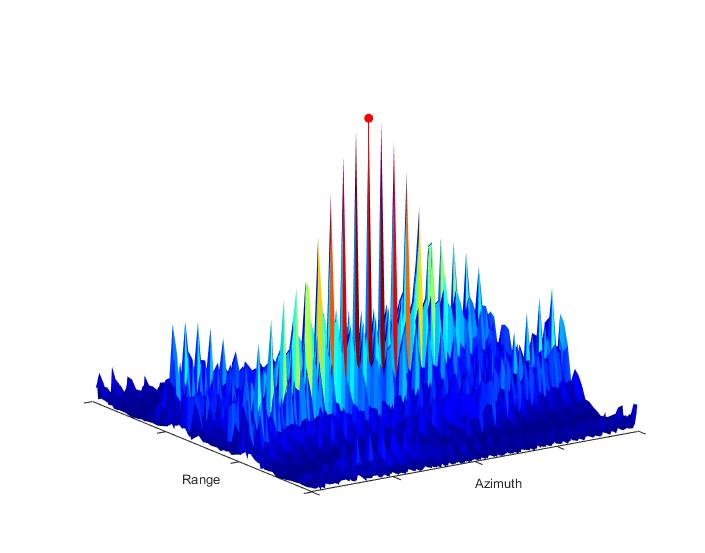}
\caption{Range-azimuth map in noiseless settings for antennas located on the conventional ULA and carrier frequencies selected on a grid, with $L=1$ target. The highest peak with the red circle corresponds to the true target. The other peaks results from the range-azimuth coupling.}
\label{fig:amb1}
\end{center}
\end{figure}

In the single pulse case, where $P=1$, we can rewrite (\ref{coffAlligned2_simp}) with respect to the channel $\gamma=mR+q$ as
\begin{equation} \label{eq:amb}
y_{\gamma}[k]=   \sum_{l=1}^{L} a_l e^{j2\pi \beta_{\gamma} \vartheta_l} e^{-j2\pi f_{\gamma} \tau_l} e^{-j\frac{2\pi }{\tau}k\tau_l},
\end{equation}
where $\beta_{\gamma}=\gamma/2$, $f_{\gamma} = (\gamma \text{ mod } T)B_h$ and $a_l$ is equal to $\alpha_l$ up to constant phases. Assume that $\tau_l$ and $\vartheta_l$ lie on the Nyquist grid such that
\begin{eqnarray} \label{grid1}
\tau_l &=& \frac{\tau}{TN}s_l, \nonumber 
\\  \vartheta_l &=& -1+\frac{2}{TR} r_l,  
\end{eqnarray} 
where $s_l$ and $r_l$ are integers satisfying $0 \leq s_l \leq TN-1$ and $0 \leq r_l \leq TR -1$, respectively.
Then, (\ref{eq:amb}) becomes
\begin{equation}
y_{\gamma}[k]=   \sum_{l=1}^{L} a_l e^{j2\pi \frac{\gamma}{TR}(r_l-s_l R)} e^{-j\frac{2\pi }{TN}ks_l},
\end{equation}
for $-\frac{N}{2} \leq k \leq \frac{N}{2}-1$. We observe that the first term is ambiguous with respect to the range and azimuth and therefore the range-azimuth beamforming is ambiguous. In noiseless settings, we can recover the range and azimuth with no ambiguity from the second term (single channel). This is illustrated in Fig.\hbox{~\ref{fig:amb1}}, where the highest peak in the range-azimuth map corresponds to the true target. However, this configuration may lead to ambiguity between both parameters in the presence of noise, due to other high peaks. We thus choose to adopt a random array, as discussed in Section~\ref{sec:sim}.

\section{Range-Azimuth-Doppler Recovery}
\label{sec:recovery}

We now describe our recovery approach from the Fourier coefficients of the FDMA received waveforms (\ref{eq:rec_sig}). We first consider the case where $P=1$ and derive range-azimuth recovery from the coefficients (\ref{coffAlligned2_simp}). We next turn to range-azimuth-Doppler recovery from (\ref{coffAlligned2}).

\subsection{Range-Azimuth Recovery}
\label{sec:algo_r_a}

In practice, as in traditional MIMO, suppose we now limit ourselves to the Nyquist grid with respect to the total bandwidth $TB_h$ so that $\tau_l$ and $\vartheta_l$ lie on the grid defined in (\ref{grid1}). Let $\mathbf{Y}^m$ be the $N \times R$ matrix with $q$th column given by $y_{m,q}[k-N/2]$, defined in (\ref{coffAlligned2_simp}), for $0 \leq k \leq N-1$. We can write $\mathbf{Y}^m$ as
\begin{equation}
\label{eq:model}
\mathbf{Y}^m = \mathbf{A}^m \mathbf{X} \left(\mathbf{B}^m\right)^T.
\end{equation}
Here, $\mathbf{A}^m$ denotes the $N \times TN$ matrix whose $(k,n)$th element is $e^{- j \frac{2 \pi}{TN} (k-\frac{N}{2}) n} e^{-j2\pi \frac{f_m}{B_h} \frac{n}{T}}$ and $\mathbf{B}^m$ is the $R \times TR$ matrix with $(q,p)$th element $e^{j2 \pi \beta_{mq} (-1 +\frac{2}{TR}p)}$. The matrix $\bf X$ is a $TN \times TR$ sparse matrix that contains the values $\alpha_l$ at the $L$ indices $\left( s_l, r_l \right)$.

Our goal is to recover $\bf X$ from the measurement matrices $\mathbf{Y}^m, 0 \leq m \leq M-1$. The time and spatial resolution induced by $\bf X$ are $\frac{\tau}{TN}=\frac{1}{TB_h}$, and $\frac{2}{TR}$, respectively, as in classic CDMA processing.

Define
\begin{equation} \label{eq:bigA}
\mathbf{A} = [ \mathbf{A}^{0^T} \, \mathbf{A}^{1^T} \, \cdots \, \mathbf{A}^{(T-1)^T}]^T,
\end{equation}
and 
\begin{equation} \label{eq:bigB}
\mathbf{B} = [ \mathbf{B}^{0^T} \, \mathbf{B}^{1^T} \, \cdots \, \mathbf{B}^{(T-1)^T} ]^T.
\end{equation}
  To better grasp the structure of $\bf A$ and $\bf B$, suppose that the carriers $f_m$ lie on the grid $f_m=(m - \frac{T-1}{2})B_h$. In this case, the $(k,n)$th element of $\mathbf{A}^m$ is $e^{- j \frac{2 \pi}{TN} (k+mN-\frac{TN}{2})n}$ and $\bf A$ is the $TN \times TN$ Fourier matrix up to row permutation. Similarly, assuming that the antenna elements lie on the virtual array illustrated in Fig.~\ref{fig:arrays1}, we have $\beta_{mq}=\frac{1}{2} (q + m R)$, where we used \textbf{A5} to simplify the expression. Then, the $(q,p)$th element of $\mathbf{B}^m$ is $e^{ j \frac{2\pi}{TR} (q+mR)(p-\frac{TR}{2})}$ and $\bf B$ is the $TR \times TR$ Fourier matrix up to column permutation. The matrices $\bf A$ and $\bf B$ are sometimes referred to as dictionaries, whose columns correspond to the range and azimuth grid points, respectively. However, this configuration leads to range-azimuth coupling as discussed in Section \hbox{\ref{sec:coupl}}. In Section~\hbox{\ref{sec:sim}}, we use a random array to avoid range-azimuth coupling.

One approach to solving \hbox{(\ref{eq:model})} is based on CS \hbox{\cite{CSBook, SamplingBook}} techniques that exploits the sparsity of the target scene. One of CS recovery advantages is that it allows to reduce the number of required samples, pulses and channels while preserving the underlying resolution \hbox{\cite{cohen2016summer}}. In particular, we adopt an iterative reconstruction approach that is beneficial when dealing with high dynamic range with both weak and strong targets, especially since the sidelobes are slightly raised due to the random array configuration. Our recovery algorithm is based on orthogonal matching pursuit (OMP) \cite{CSBook, SamplingBook}. Similar subtraction techniques are used in many iterative algorithms such as the CLEAN process \hbox{\cite{tsao1988reduction}}.

To recover the sparse matrix $\bf X$ from the set of equations (\ref{eq:model}), for all $0 \leq m \leq M-1$, where the targets' range and azimuth lie on the Nyquist grid, we consider the following optimization problem
\begin{equation}
\label{eq:opt}
\text{min } ||\mathbf{X}||_0 \text{ s.t. } \mathbf{A}^m \mathbf{X} \left( \mathbf{B}^m\right)^T=\mathbf{Y}^m, \quad 0 \leq m \leq T-1.
\end{equation}
It has been shown in \hbox{\cite{cohen2016summer}} that the minimal number of channels required for perfect recovery of $\mathbf{X}$ in (\ref{eq:opt}) with $L$ targets in noiseless settings is $TR \geq 2L$ with a minimal number of $TN \geq 2L$ samples per receiver.
To solve (\ref{eq:opt}), we extend the matrix OMP from \cite{cs_mat_yonina} to simultaneously solve a system of CS matrix equations, as shown in Algorithm \ref{algo:omp}. In the algorithm description, $ \text{vec}(\mathbf{Y}) \triangleq \left[ \text{vec}(\mathbf{Y}^0)^T \, \cdots \,      \text{vec}(\mathbf{Y}^{T-1})^T  \right]^T$, $\mathbf{d}_t(l)= \left[ (\mathbf{d}_t^0(l))^T \, \cdots \, (\mathbf{d}_t^{T-1}(l))^T \right]^T$ where $\mathbf{d}_t^m(l) = \text{vec}(\mathbf{a}^m_{\Lambda_t(l,1)} (\mathbf{b}^m_{\Lambda_t(l,2)})^T)$ with $\Lambda_t(l,i)$ the $(l,i)$th element in the index set $\Lambda_t$ at the $t$th iteration, and $\mathbf{D}_t= [\mathbf{d}_t(1) \, \dots \, \mathbf{d}_t(t)]$. Here, $\mathbf{a}^m_j$ denotes the $j$th column of the matrix $\mathbf{A}^m$ and $\mathbf{b}^m_j$ denotes the $j$th column of the matrix $\mathbf{B}^m$. Once $\bf X$ is recovered, the delays and azimuths are estimated as
\begin{eqnarray}
\hat{\tau}_l &=& \frac{\tau}{TN}\Lambda_L(l,1), \label{eq:hat_tau} \\ \quad \hat{\vartheta}_l &=& -1+\frac{2}{TR}\Lambda_L(l,2). \label{eq:hat_theta}
\end{eqnarray}
\begin{algorithm}
	\caption{Simultaneous sparse 2D recovery based OMP}\label{algo:omp} 
	\begin{algorithmic}[1]
		\qinput Observation matrices $\mathbf{Y}^m$, measurement matrices $\mathbf{A}^m$, $\mathbf{B}^m$, for all $0 \leq m \leq T-1$
		\qoutput Index set $\Lambda$ containing the locations of the non zero indices of $\mathbf{X}$, estimate for sparse matrix $\bf \hat{X}$
		\State Initialization: residual $\mathbf{R}_0^m=\mathbf{Y}^m$, index set $\Lambda_0=\emptyset$, $t=1$
		\State Project residual onto measurement matrices:
		$$
		\mathbf{\Psi} =\mathbf{A}^H \mathbf{R} \mathbf{\bar{B}}
		$$	
		where $\mathbf{A}$ and $\mathbf{B}$ are defined in (\ref{eq:bigA}) and (\ref{eq:bigB}), respectively, and $\mathbf{R} = \text{diag} \left( [ \mathbf{R}^0_{t-1} \, \cdots \, \mathbf{R}^{T-1}_{t-1}] \right)$ is block diagonal
		\State Find the two indices $\lambda_t = [\lambda_t(1) \quad \lambda_t(2)]$ such that
		$$
		[\lambda_t(1) \quad \lambda_t(2)] = \text{ arg max}_{i,j} \left| \mathbf{\Psi}_{i,j} \right|
		$$
		\State Augment index set $\Lambda_t = \Lambda_t  \bigcup \{ \lambda_t \}$
		\State Find the new signal estimate
		$$
		\mathbf{\hat{\alpha}}=[\hat{\alpha}_1 \, \dots \, \hat{\alpha}_t]^T = ( \mathbf{D}_t^T \mathbf{D}_t )^{-1} \mathbf{D}_t^T \text{vec}(\mathbf{Y}) 
		$$
		\State Compute new residual
		$$
		\mathbf{R}_t^m= \mathbf{R}_0^m- \sum_{l=1}^t \alpha_l \mathbf{a}^m_{\Lambda_t(l,1)} \left( \mathbf{b}^m_{\Lambda_t(l,2)} \right)^T$$
		\State If $t < L$, increment $t$ and return to step 2, otherwise stop
		\State Estimated support set $\hat{\Lambda}= \Lambda_L$ \\
		Estimated matrix $\bf \hat{X}$: $\left( \Lambda_L(l,1), \Lambda_L(l,2) \right)$-th component of $\bf \hat{X}$ is given by $\hat{\alpha}_l$ for $l=1, \cdots, L$ while rest of the elements are zero
	\end{algorithmic}
\end{algorithm}
Other CS recovery algorithms, such as FISTA \cite{FISTA_beck, OptiBook}, can also be extended to our setting.

The projection performed in step 2 of the algorithm combines single channel processing with range-azimuth beamforming. The former coherently processes the second term of \hbox{(\ref{coffAlligned2_simp})}, which appears in the matrix $\bf A$, while range-azimuth beamforming over the channels coherently processes the first and third terms of \hbox{(\ref{coffAlligned2_simp})}, which are contained in $\bf A$ and $\bf B$, respectively. The FDMA narrowband assumption reconciliation is due to the additional third term of \hbox{(\ref{eq:term1_1})}, contained in $\bf B$,  thus enhancing range-azimuth resolution capabilities.

The improved performance of the iterative approach over non-iterative target recovery with high dynamic range is illustrated in simulations in Section~\ref{sec:sim}. There, we also compare our FDMA approach with classic CDMA, when using a non-iterative recovery method in the former. In particular, we only use one iteration of Algorithm \ref{algo:omp}, which is equivalent to the classic approach. This demonstrates that our FDMA method outperforms CDMA due to the high range-azimuth resolution capabilities stemming from the reconciliation between the individual narrowband assumption and the large overall bandwidth.

\subsection{Range-Azimuth-Doppler Recovery}
\label{sec:algo}

Besides $\tau_l$ and $\vartheta_l$ lying on the grid defined in (\ref{grid1}), we assume that the Doppler frequency $f^D_l$ is limited to the Nyquist grid as well, defined by the CPI as:
\begin{equation} \label{grid2}
f^D_l = -\frac{1}{2\tau}+\frac{1}{P\tau}u_l,
\end{equation} 
where $u_l$ is an integer satisfying $0 \leq u_l \leq P-1$. Let $\mathbf{Z}^m$ be the $NR \times P$ matrix with $q$th column given by the vertical concatenation of $y_{m,q}^p[k]$, such that the $(k+qN,p)$th element of $\mathbf{Z}^m$ is given by $(\mathbf{Z}^m)_{(k+qN,p)}=y_{m,q}^p[k-N/2]$, defined in (\ref{coffAlligned2}), for $0 \leq k \leq N-1$ and $0 \leq q \leq R-1$. We can then write $\mathbf{Z}^m$ as

\begin{equation}
\label{eq:model2}
\mathbf{Z}^m = \left( \mathbf{B}^m \otimes  \mathbf{A}^m \right) \mathbf{X}_D \mathbf{F}^T,
\end{equation}
where the $N \times TN$ matrix $\mathbf{A}^m$ and the $R \times TR$ matrix $\mathbf{B}^m$ are defined as in Section \hbox{\ref{sec:algo_r_a}} and $\bf F$ denotes the $P \times P$ Fourier matrix up to column permutation. The matrix $\mathbf{X}_D$ is a $T^2NR \times P$ sparse matrix that contains the values $\alpha_l$ at the $L$ indices $\left( r_l TN +s_l, u_l \right)$.

Our goal is now to recover $\mathbf{X}_D$ from the measurement matrices $\mathbf{Z}^m, 0 \leq m \leq T-1$. The time, spatial and frequency resolution stipulated by $\mathbf{X}_D$ are $\frac{\tau}{TN}=\frac{1}{B_{\text{tot}}}$ with $B_{\text{tot}}=TB_h$, $\frac{2}{TR}$ and $\frac{1}{P \tau}$ respectively, as in classic CDMA processing.

To jointly recover the range, azimuth and Doppler frequency of the targets, we apply the concept of Doppler focusing from \cite{bar2014sub} to our setting. Once the Fourier coefficients (\ref{coffAlligned2}) are processed, we perform Doppler focusing for a specific frequency $\nu$, that is
    \begin{eqnarray}
\label{eq:dop_reduced}
   \Phi^{\nu}_{m,q}[k] &=&\sum_{p=0}^{P-1} y_{m,q}^p[k] e^{j2\pi \nu p \tau} \\
 &=&   \sum_{l=1}^{L} \alpha_l e^{j2\pi \beta_{mq} \vartheta_l}e^{-j\frac{2\pi }{\tau}(k+f_m \tau)\tau_l} \sum_{p=0}^{P-1} e^{j 2 \pi (\nu-f^D_l) p\tau}, \nonumber
       \end{eqnarray}
       for $-\frac{N}{2} \leq k \leq -\frac{N}{2}-1$.
Following the same argument as in \cite{bar2014sub}, it holds that
\begin{equation}
\label{eq:dop_foc}
\sum_{p=0}^{P-1} e^{j 2 \pi (\nu-f^D_l) p\tau} \cong \left\{ \begin{array}{ll} P & |\nu-f^D_l| < \frac{1}{2P\tau}, \\
0 & \text{otherwise}. \end{array} \right.
\end{equation}
Therefore, for each focused frequency $\nu$,  (\ref{eq:dop_reduced}) reduces to a 2-dimensional problem. We note that Doppler focusing increases the SNR by a factor a $P$, as can be seen in (\ref{eq:dop_foc}).

Algorithm \ref{algo:omp2} solves (\ref{eq:model2}) for $0 \leq m \leq T-1$ using Doppler focusing. Note that step 1 can be performed using the fast Fourier transform (FFT). In the algorithm description, $\text{vec}(\mathbf{Z})$ is defined as in the previous section, $\mathbf{e}_t(l)= \left[ (\mathbf{e}_t^0(l))^T \, \cdots \, (\mathbf{e}_t^{T-1}(l))^T \right]^T$ where $\mathbf{e}_t^m(l) = \text{vec}( (\mathbf{B}^m \otimes \mathbf{A}^m)_{\Lambda_t(l,2)TN+\Lambda_t(l,1)} (\mathbf{F}^m_{\Lambda_t(l,3)})^T)$ with $\Lambda_t(l,i)$ the $(l,i)$th element in the index set $\Lambda_t$ at the $t$th iteration, and $\mathbf{E}_t= [\mathbf{e}_t(1) \, \dots \, \mathbf{e}_t(t)]$.
Once $\mathbf{X}_D$ is recovered, the delays and azimuths are given by (\ref{eq:hat_tau}) and (\ref{eq:hat_theta}), respectively, and the Doppler frequencies are estimated as
\begin{equation}
\label{eq:hat_f}
\hat{f}_l^D = -\frac{1}{2\tau}+\frac{\Lambda_L(l,3)}{P\tau}.
\end{equation}
Similarly to the one-pulse case, the minimal number of channels required for perfect recovery of $\mathbf{X}_D$ with $L$ targets in noiseless settings is $TR \geq 2L$ with a minimal number of $TN \geq 2L$ samples per receiver and $P \geq 2L$ pulses per transmitter \hbox{\cite{cohen2016summer}}.

\begin{algorithm}
\caption{Simultaneous sparse 3D recovery based OMP with focusing}\label{algo:omp2} 
\begin{algorithmic}[1]
\qinput Observation matrices $\mathbf{Z}^{m}$, measurement matrices $\mathbf{A}^{m}$, $\mathbf{B}^{m}$, for all $0 \leq m \leq T-1$
\qoutput Index set $\Lambda$ containing the locations of the non zero indices of $\mathbf{X}_D$, estimate for sparse matrix $\mathbf{\hat{X}}_D$
\State Perform Doppler focusing for $0 \leq i \leq N-1$, $0 \leq j \leq R-1$ and $0 \leq \nu \leq P-1$ :
$$
\mathbf{\Phi}^{(m,\nu)}_{i,j} = (\mathbf{Z}^m \mathbf{\bar{F}})_{i+jN,\nu}.
$$
\State Initialization: residual $\mathbf{R}_0^{(m,\nu)}=\mathbf{\Phi}^{(m,\nu)}$, index set $\Lambda_0=\emptyset$, $t=1$
\State Project residual onto measurement matrices for $0 \leq \nu \leq P-1$:
$$
\mathbf{\Psi}^\nu =\mathbf{A}^H \mathbf{R}^{\nu} \mathbf{\bar{B}},
$$
where $\mathbf{A}$ and $\mathbf{B}$ are defined in (\ref{eq:bigA}) and (\ref{eq:bigB}), respectively, and $\mathbf{R}^{\nu} = \text{diag} \left( [ \mathbf{R}^{(0,\nu)}_{t-1} \, \cdots \, \mathbf{R}^{(T-1, \nu)}_{t-1}] \right)$ is block diagonal
\State Find the three indices $\lambda_t = [\lambda_t(1) \, \lambda_t(2) \, \lambda_t(3)]$ such that
$$
[\lambda_t(1) \quad \lambda_t(2) \quad \lambda_t(3)] = \text{ arg max}_{i,j,\nu} \left| \mathbf{\Psi}^{\nu}_{i,j} \right|
$$
\State Augment index set $\Lambda_t = \Lambda_t  \bigcup \{ \lambda_t \}$
\State Find the new signal estimate
$$
\mathbf{\hat{\alpha}}=[\hat{\alpha}_{1} \, \dots \, \hat{\alpha}_{t}]^T = ( \mathbf{E}_t^T \mathbf{E}_t )^{-1} \mathbf{E}_t^T \text{vec}(\mathbf{Z}) 
$$
\State Compute new residual
$$
\mathbf{R}_t^{(m,\nu)}= \mathbf{R}_0^{(m,\nu)}- \sum_{l=1}^t \alpha_l \mathbf{a}^m_{\Lambda_t(l,1)} \left( \mathbf{b}^m_{\Lambda_t(l,2)} \right)^T\left( \mathbf{f}_{\Lambda_t(l,3)} \right)^T\mathbf{f}_{\nu}$$
\State If  $t < L$, increment $t$ and return to step 3, otherwise stop
\State Estimated support set $\hat{\Lambda}= \Lambda_L$ \\
Estimated matrix $\mathbf{\hat{X}}_D$: $\left( \Lambda_L(l,2) TN+ \Lambda_L(l,1) ,  \Lambda_L(l,3)\right)$-th component of $\mathbf{\hat{X}}_D$ is given by $\hat{\alpha}_l$ for $l=1, \cdots, L$ while rest of the elements are zero
\end{algorithmic}
\end{algorithm}

\section{Simulations}
\label{sec:sim}

In this section, we present numerical experiments illustrating our FDMA approach and compare our method with classic MIMO processing using CDMA.

\subsection{Preliminaries}

Throughout the experiments, the standard MIMO system is based on a virtual array, as depicted in Fig.~\ref{fig:arrays1} generated by $T=20$ transmit antennas and $R=20$ receive antennas, yielding an aperture $\lambda Z=6m$. We consider a random array configuration where the transmitters and receivers' locations are selected uniformly at random over the aperture $Z$. We use FDMA waveforms $h_m(t)$ such that $f_m=(i_m-\frac{T-1}{2})B_h$, where $i_m$ are integers chosen uniformly at random in $[0,T)$, for $0 \leq m \leq T-1$, and all frequency bands within $[-\frac{T}{2}B_h, \frac{T}{2}B_h]$ are used for transmission. We consider the following parameters: PRI $\tau=100 \mu sec$, bandwidth $B_h = 5MHz$ and carrier frequency $f_c=10 GHz$. We simulate targets from the Swerling-0 model with identical amplitudes and random phases. The received signals are corrupted by uncorrelated additive Gaussian noise (AWGN) with power spectral density $N_0$. The SNR is defined as
\begin{equation}
\text{SNR} = \frac{\frac{1}{T_p} \int_0^{T_p} |h_0(t)|^2 \mathrm{d}t}{N_0  B_h},
\end{equation}
where $T_p$ is the pulse time.

We consider a hit-or-miss criterion as performance metric. A ``hit" is defined as a range-azimuth estimate which is identical to the true target position up to one Nyquist bin (grid point) defined as $1/TB_h$ and $2/TR$ for the range and azimuth, respectively. In pulse-Doppler settings, a ``hit" is proclaimed if the recovered Doppler is identical to the true frequency up to one Nyquist bin of size $1/P\tau$, in addition to the two previous conditions.
\subsection{Numerical Results}

\begin{figure}
\begin{center}
\includegraphics[width=0.5\textwidth]{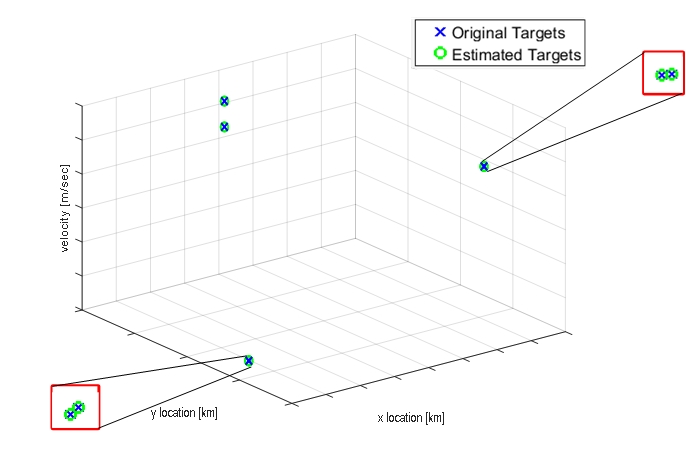}
\caption{Range-azimuth-Doppler recovery for $L=6$ targets and SNR=$-10$dB.}
\label{fig:close}
\end{center}
\end{figure}

We first consider a sparse target scene with $L=6$ targets including a couple of targets with close ranges, a couple with close azimuths and another couple with close velocities. We use $P=10$ pulses and the SNR is set to $-10$dB. As can be seen in Fig.~\ref{fig:close}, all targets are perfectly recovered, demonstrating high resolution in all dimensions. Here, the range and azimuth are converted to 2-dimensional $x$ and $y$ locations. 

\begin{figure}
\begin{center}
\includegraphics[width=0.5\textwidth]{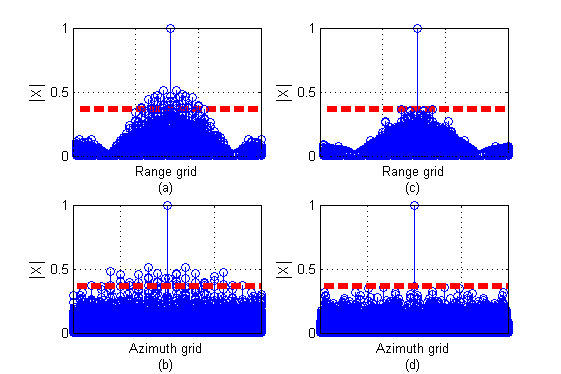}
\caption{Range-azimuth map in noiseless settings for random carrier frequencies along range axis (a) and azimuth axis (b), and for random antennas' locations along range axis (c) and azimuth axis (d), for $L=1$ target. The red dotted line indicates the peak sidelobe level for this target.}
\label{fig:amb2}
\end{center}
\end{figure}

We next turn to the range-azimuth coupling issue and discuss the impact of the choice of antennas' locations and transmissions' carrier frequencies. As discussed in Section~\hbox{\ref{sec:coupl}}, the conventional ULA array structure shown in Fig.~\ref{fig:arrays1} with carrier frequency selected on a grid, leads to ambiguity in the range-azimuth domain. In order to overcome the ambiguity issue, we adopt a random array configuration \cite{rossi2014spatial}. We found heuristically that a configuration with random antennas' locations with carriers on a grid provides better results than random carriers with a ULA structure. Figure~\ref{fig:amb2} shows a typical result of sidelobes for both configurations. The peak sidelobe level for the configuration with random antennas' locations is consistently lower.

\begin{figure}
\begin{center}
\includegraphics[width=0.5\textwidth]{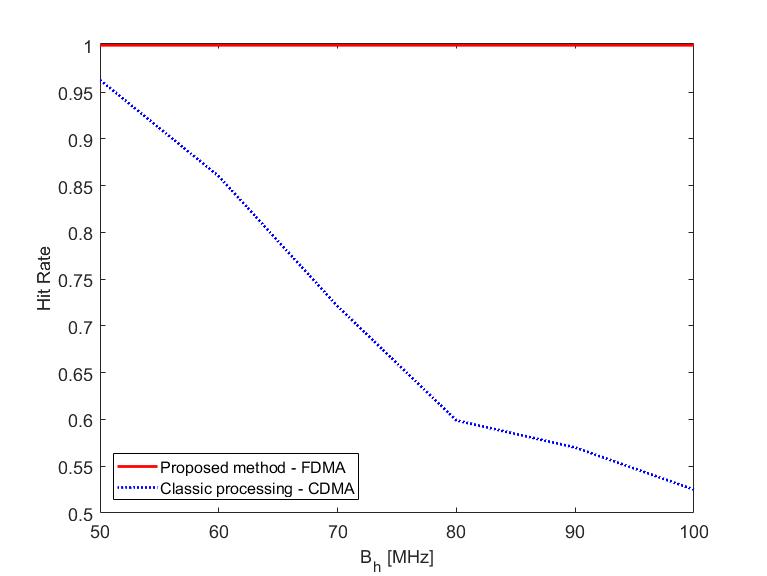}
\caption{Hit rate of FDMA and classic CDMA versus bandwidth.}
\label{fig:fdma_cdma}
\end{center}
\end{figure}

\begin{figure}
\begin{center}
\includegraphics[width=0.5\textwidth]{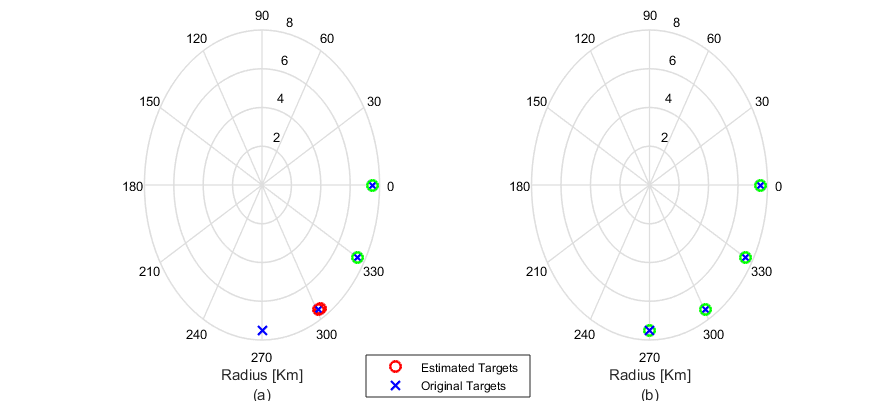}
\caption{Range-azimuth recovery for $L=4$ targets using classic CDMA (a) and FDMA (b) .}
\label{fig:fdma_cdma_t}
\end{center}
\end{figure}

We then compare our FDMA processing with classic MIMO processing using CDMA waveforms. Figure~\ref{fig:fdma_cdma} shows the hit rate of both techniques with respect to bandwidth $B_h$ so that the total bandwidth $B_{\text{tot}}$ is identical for both. The experiment was performed without noise so that the decrease in performance in CDMA is due to the violation of the narrowband array assumption (\ref{eq:A1}). When the received signal is modeled such as to artificially remove the delay differences between antennas, that is synthetically generated so that $\tau_l$ replaces $\tau_{l,mq}$ in (\ref{eq:term3}), then the performance of both methods is identical, as expected. In Fig.~\ref{fig:fdma_cdma_t}, we observe that the targets with small azimuth angle $\theta_l$ are detected by both techniques, whereas targets on the end-fire direction ($\theta_l=\pm 90\si{\degree}$, corresponds to the broadside direction) are missed by the CDMA approach. This happens because the delay differences between channels are too large, which violates the narrowband assumption \textbf{A5}.

\begin{figure}
\begin{center}
\includegraphics[width=0.5\textwidth]{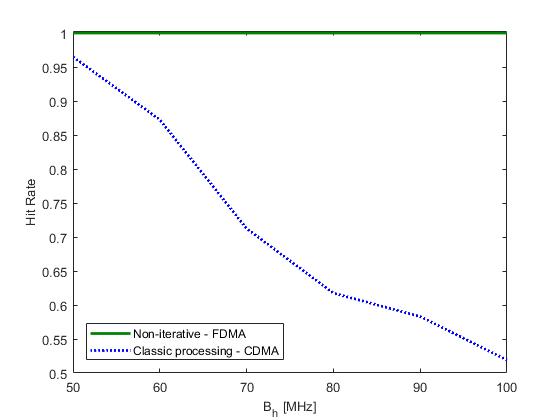}
\caption{Hit rate of non-iterative FDMA and classic CDMA versus bandwidth.}
\label{fig:nonIterative_fdma_cdma}
\end{center}
\end{figure}

In order to demonstrate that the performance gain of our FDMA approach over the classic CDMA is due to the relaxed narrowband assumption rather than our specific iterative processing, we consider a non-iterative recovery approach. \hbox{Figure~\ref{fig:nonIterative_fdma_cdma}} shows our FDMA method with non-iterative recovery, corresponding to one iteration of Algorithm~\ref{algo:omp}. This constitutes evidence that our approach outperforms conventional CDMA processing from the relaxed narrowband assumption. 

\begin{figure}
\begin{center}
\includegraphics[width=0.5\textwidth]{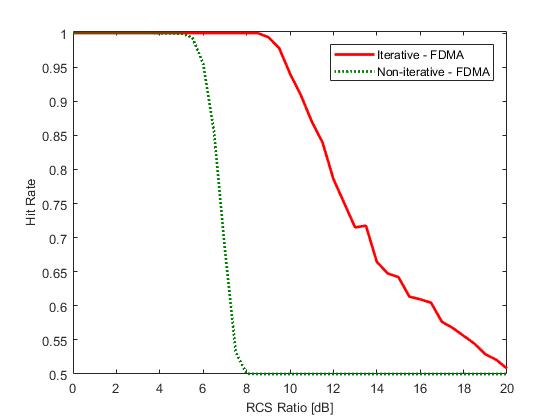}
\caption{Hit rate of iterative FDMA and non-iterative FDMA versus RCS ratio.}
\label{fig:rcs_dinamicrange}
\end{center}
\end{figure}

The iterative approach does boost performance of multiple targets recovery with high RCS dynamic range, allowing detection of weak targets masked by the strong ones. In doing so, we further decrease the effect of the sidelobe level and thus improve detection performance. To compare both iterative and non iterative recovery, we consider $L=2$ targets whose locations are generated uniformly at random with varying RCS ratios defined as $10\log_{10}\left(\frac{\alpha_{l_{max}}}{\alpha_{l_{min}}}\right)$. In \hbox{Fig.~\ref{fig:rcs_dinamicrange}}, we can see that the non-iterative approach attains $50\%$ hit rate for an RCS ratio of 8 dB, which means that the weak target is totally masked by the strong one. The iterative approach detects the weak target up to an RCS ratio of 20 dB.

\begin{figure}
\begin{center}
\includegraphics[width=0.5\textwidth]{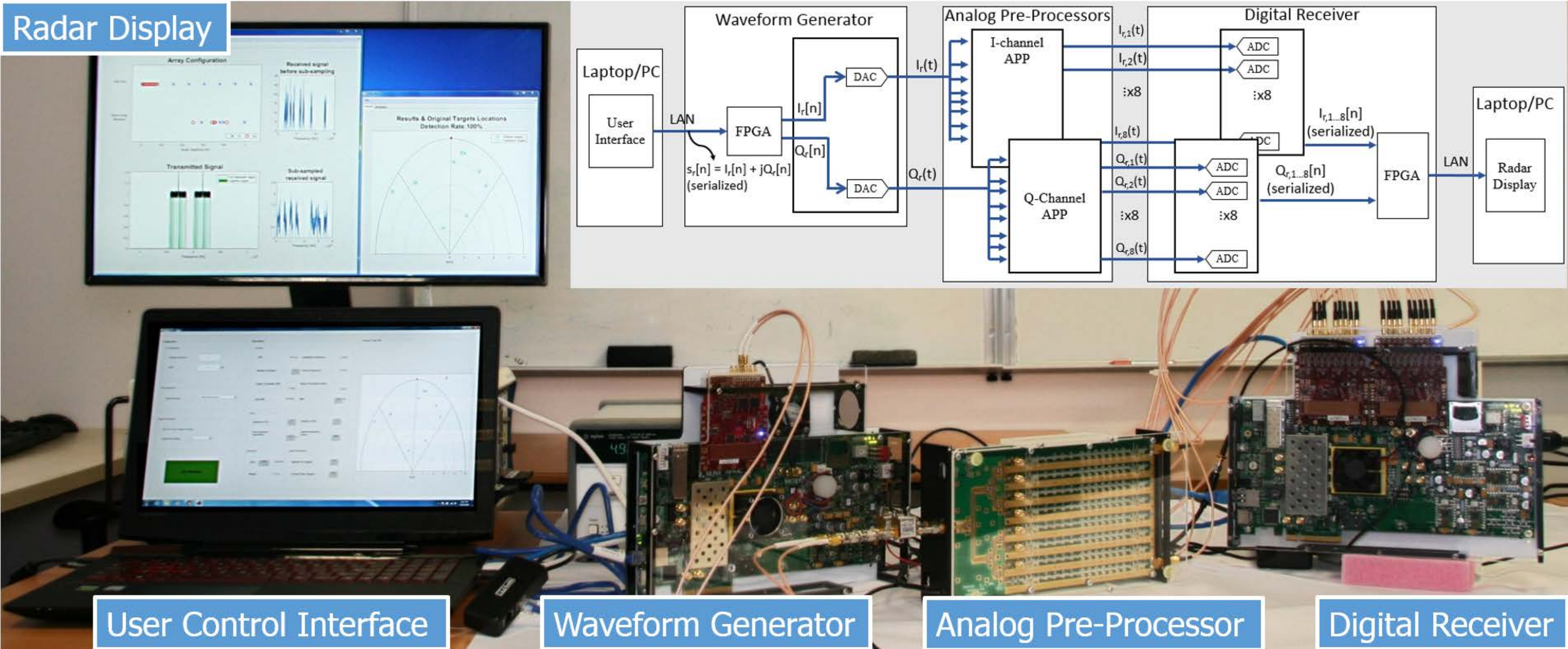}
\caption{FDMA MIMO prototype and user interface~\cite{cohen2017sub}.}
\label{fig:blockandphoto}	
\end{center}
\end{figure}

Each iteration of our proposed FDMA approach takes 3.9 sec for 40 million range-azimuth-Doppler grid points using an Intel Core i7 PC without GPU components. We have implemented a hardware prototype realizing the FDMA MIMO processing presented here. 
The prototype, shown in \hbox{Fig.~\ref{fig:blockandphoto}}, proves the hardware feasibility of our FDMA MIMO radar. Further details can be found in \hbox{\cite{mishra2016cognitive}, \cite{cohen2017sub}.}

\section{Conclusion}
In this work, we considered a MIMO radar configuration based on FDMA waveforms. Using FDMA allows us to relax the traditional narrowband assumption that creates a trade-off between range and azimuth resolution. We are able to combine large overall bandwidth for high range resolution and small individual bandwidth for high azimuth resolution. In order to overcome one of the main FDMA's drawbacks, that limits the range resolution to the individual bandwidth, we proposed a joint processing algorithm of the channels achieving range resolution with respect to the overall bandwidth. A large virtual array aperture, that yields high azimuth resolution, is enabled by the relaxed narrowband assumption and appropriate digital processing. Our system and subsequent processing copes with range-azimuth coupling, which occurs when using FDMA, by using a random array configuration. The digital processing is a feasible iterative CS based approach for simultaneous sparse recovery. Simulations illustrated the increased resolution obtained by our approach in comparison with classic CDMA, leading to better detection performance.

\bibliographystyle{IEEEtran}
\bibliography{Bib}

\end{document}